\documentclass[twocolumn,amsmath,amssymb]{snp}
\usepackage{graphicx}
\usepackage{dcolumn}
\usepackage{bm}
\newcommand{\be}{\begin{equation}}
\newcommand{\ee}{\end{equation}}
\newcommand{\bea}{\begin{eqnarray}}
\newcommand{\eea}{\end{eqnarray}}

\newcommand{\ep}{\epsilon}

\newcommand{\bp}{\boldsymbol{p}}

\begin{document}

\title{{\Large  Causal aspects of effective QCD models}}

\author{\large Jayanta Dey}
\email{jayantad@iitbhilai.ac.in}
\affiliation{Indian Institute of Technology Bhilai, GEC Campus, Sejbahar, Raipur 492015, 
Chhattisgarh, India}
\author{Chowdhury Aminul Islam}
\affiliation{School of Nuclear Science and Technology, University of Chinese Academy of Sciences, Beijing 100049, China.}
\affiliation{Department of theoretical Physics, Tata Institute of fundamental Research, 
Homi Bhabha Road, Mumbai 400005, India}
\author{Sabyasachi Ghosh}
\affiliation{Indian Institute of Technology Bhilai, GEC Campus, Sejbahar, Raipur 492015, Chhattisgarh, India}

\maketitle
Let us start with energy momentum tensor of relativistic fluid,
\be
T^{\mu\nu}=T^{\mu\nu}_0 +\pi^{\mu\nu},
\ee
where ideal part is $T^{\mu \nu}_0=\epsilon u^{\mu}u^{\nu}-P \Delta^{\mu \nu}$
and shear dissipation part is~\cite{JD_EPNJL}
\be
\pi^{\mu \nu}=\eta {\cal U}^{\mu\nu}_\eta~, 
\label{N-S}
\ee
commonly known as Navier-Stokes equation.  
If the fluid is quark matter, having $N_f$ flavor and $N_c$ color
degeneracy factors, then from kinetic theory framework,
energy-momentum tensor can be define as (for zero quark chemical potential)
\be
T^{\mu\nu}=4N_fN_c\int\frac{d^3\bp}{(2\pi)^3}\frac{p^\mu p^\nu}{E}f_{Q}~,
\label{eq:Tmunu}
\ee 
where a small deviation $\delta f_{Q}$ of distribution function $f_{Q}$ from 
equilibrium distribution $f^0_{Q}$ will be assumed, {\it i.e.}
\be
f_{Q}=f^0_{Q}+\delta f_{Q}~.
\ee
By relating macroscopic Eq.~(\ref{N-S}) to microscopic Eq.~(\ref{eq:Tmunu}) and
identifying proper expression of $\delta f_{Q}$ through relaxation time
approximation (RTA) of relativistic Boltzmann equation (RBE), we will get
\be
\eta = \frac{4N_FN_c}{15T}
\int \frac{d^3\bp}{(2\pi)^3}\left(\frac{\bp^2}{E}\right)^2\tau_{Q}{f_Q}  (1-f_Q)
\label{eta}
\ee

Now, Eq.~\ref{N-S} violates causality. 
To check it, one can derive diffusion speed $v_T$ from dispersion relation, based 
on Eq.~(\ref{N-S})~\cite{Romatchke:2009},
\be
v_T(k) = 2\frac{\eta}{\epsilon + P}k 
\ee
which can exceed speed of light at $k\rightarrow\infty$ with 
finite values of energy density $\ep$ and pressure $P$. 
\begin{figure}[hbt]
	\begin{center}
	\includegraphics[scale=0.8]{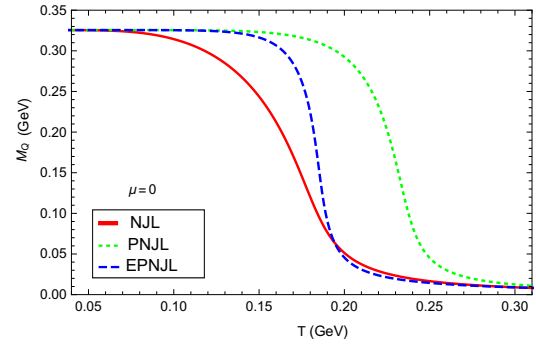}
	\end{center}
	\caption{Temperature dependence of constituent quark masses for 
		NJL (solid line), PNJL (dotted line), EPNJL (dash line).}
	\label{fig:mass2}
\end{figure}
\begin{figure} [ht]
	\includegraphics[scale=0.8]{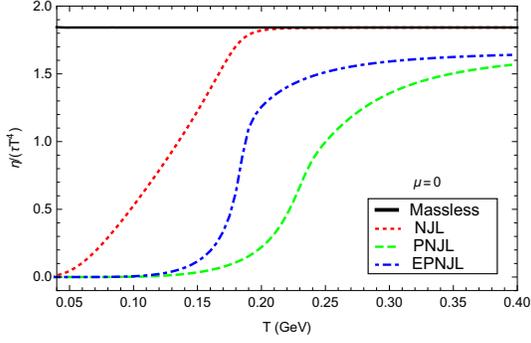}
	\caption{Normalized shear viscosity $\eta/(\tau_c T^4)$ vs temperature $T$ for massless quarks 
	(black solid horizontal line), massive quarks, based on NJL (dotted line), EPNJL (dash-dotted line), 
	PNJL (dash line).}
	\label{fig:ShT4}
\end{figure}
\begin{figure} [ht]
	\includegraphics[scale=0.8]{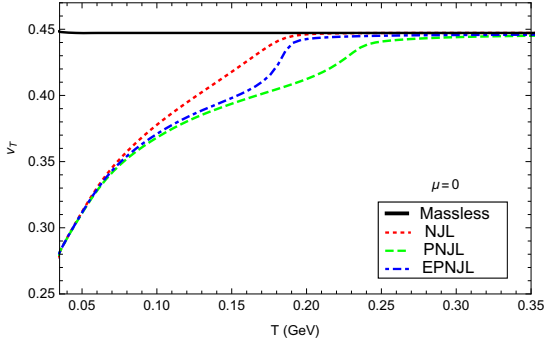}
	\caption{Maximum value of causal diffusion speed $v_T$ as a 
	function of temperature for massless quarks (black solid horizontal line), 
	massive quarks, based on NJL (dotted line), EPNJL (dash-dotted line), PNJL (dash line).}
	\label{fig:Tau_KSS}
\end{figure}

It was Maxwell~\cite{Max} and Cattaneo~\cite{Cattaneo}, who replaced the algebraic
Eq.~(\ref{N-S}) by time derivative equation, where causality can be maintained
by introducing a finite values of shear relaxation time $\tau_\pi$. In
the limit of $\tau_\pi\rightarrow 0$, Maxwell-Cattaneo equation is merged to acausal 
Eq.~(\ref{N-S}). For this Maxwell-Cattaneo equation, one get diffusion speed at $k\rightarrow\infty$,
\be
v_T^{max} \equiv \lim\limits_{k\rightarrow \infty} \sqrt{\frac{\eta}{(\epsilon + P)\tau_{\pi}}}
\label{vT_max}
\ee
Interestingly, for all known fluids the limiting value of $v_T^{max}$
has been found to be smaller than one. It means that diffusion speed of our 
known fluid never cross speed of light and causality is always preserved. In present
investigation, we will explore on estimation of $v_T^{max}$ for interacting quark matter,
where interaction picture is mapped by different effective QCD models, like
Nambu-Jona-Lasinio (NJL) model~\cite{NJL}, Polyakov loop extended NJL (PNJL) model~\cite{PNJL}
and entangled PNJL (EPNJL) model~\cite{EPNJL}.
These effective QCD models build a quasi-particle description of thermodynamics
for interacting medium, where massless quark with Fermi-Dirac distribution is
transformed to temperature dependent mass in NJL model and additional modified
distribution functions in PNJL and EPNJL models. A detail discussion on these models
are addressed in Ref.~\cite{JD_EPNJL}. Using their constituent quark masses $M_Q(T)$,
shown in Fig~(\ref{fig:mass2}), in Eq.~(\ref{eta}), we have estimated respective $\eta$'s, 
which are displayed in Fig.~(\ref{fig:ShT4}). Assuming $\tau_\pi=\tau_Q$ in Eq.~(\ref{vT_max}),
we have plotted $v_T(T)$ for different model in Fig.~(\ref{fig:Tau_KSS}). Which respect massless
limit of $\eta\propto T^4$ and $v_T=1/\sqrt{5}$, interacting values remain lower in low temperature
domain. It reflects that massless non-interacting quark matter as well as interacting massive quark
matter, based on different effective QCD models both obey causal description and investigation explore
the zone with respect to the marginal values, beyond which causality can not be preserved.

\end{document}